\documentclass[pre,twocolumn,superscriptaddress]{revtex4-1}

\usepackage{graphicx}
\usepackage{dcolumn}
\usepackage{amsmath}
\usepackage{bm}
\usepackage{enumitem}
\usepackage{algorithm}
\usepackage{algpseudocode}
\usepackage{enumitem}
\setlist{leftmargin=0.5mm}
\usepackage{xcolor}

\newcommand{\Ord}{\mathrm{O}}
\newcommand{\abs}[1]{\vert #1 \vert}

\renewcommand{\vec}{\bm}

\begin{document}

\title{Spatial regionalization based on optimal information compression}

\author{Alec Kirkley${}^\ast$}
\affiliation{Institute of Data Science, University of Hong Kong, Hong Kong, China}
\affiliation{Department of Urban Planning and Design, University of Hong Kong, Hong Kong, China}
\affiliation{Urban Systems Institute, University of Hong Kong, Hong Kong, China}

\begin{abstract}
\begin{center}
${}^\ast$ email: \url{alec.w.kirkley@gmail.com}\\
\end{center}
\begin{center}
\normalsize\textbf{Abstract}\\
\end{center}
\indent\indent Regionalization, spatially contiguous clustering, provides a means to reduce the effect of noise in sampled data and identify homogeneous areas for policy development among many other applications. Existing regionalization methods require user input such as the number of regions or a similarity measure between regions, which does not allow for the extraction of the natural regions defined solely by the data itself. Here we view the problem of regionalization as one of data compression and develop an efficient, parameter-free regionalization algorithm based on the minimum description length principle. We demonstrate that our method is capable of recovering planted spatial clusters in noisy synthetic data, and that it can meaningfully coarse-grain real demographic data. Using our description length formulation, we find that spatial ethnoracial data in U.S. metropolitan areas has become less compressible over the period from 1980 to 2010, reflecting the rising complexity of urban segregation patterns in these metros.
\end{abstract}
\maketitle


\section{Introduction}

From the growth of economies \cite{fujita1999spatial} to the systemic segregation of human populations \cite{brown2006spatial} to the environmental adaptation of ecological species \cite{legendre1989spatial}, many social and natural phenomena manifest themselves in space with high levels of clustering among similar agents or entities. Precisely defining the spatial boundaries of these clusters and observing their evolution can shed light on the fundamental processes driving the dynamics of these systems, aid in the reduction of noise in spatially sampled data \cite{spielman2015reducing,spielman2015studying}, and facilitate the identification of regions for spatially targeted policy interventions \cite{rahman2004regionalization} among numerous other applications. Regionalization methods---techniques to perform spatially constrained clustering by aggregating spatial units---are typically the tools of choice for partitioning spatial data into areas of interest for such analysis. Consequently, regionalization methods have been adapted for applications across fields as diverse as climatology \cite{fovell1993climate}, urban sociology \cite{garreton2016santiago}, hydrology \cite{peterson2011hydro}, geoecology \cite{niesterowicz2016unsupervised}, and political science \cite{george1997political}.

Many approaches to regionalization typically require a significant amount of input from the user to adjust various parameters prior to performing the clustering. These tunable parameters can be used to constrain the size or shape of clusters, or to avoid crossing administrative or geographical boundaries \cite{duque2007supervised,li2013efficient}. User preferences are also commonly incorporated into regionalization methods through the choice of a similarity or distance function between adjacent regions \cite{assunccao2006efficient,wei2021efficient}. Additionally, as is the case with any clustering method, a key factor existing regionalization methods consider is the choice of the number of regions, which is typically fixed by the user \cite{duque2007supervised,aydin2021quantitative} but is sometimes determined endogeneously based on user-defined thresholds for covariates of interest or other heuristics that depend or one's choice of dissimilarity between spatial units \cite{duque2012max,wei2021efficient}. An increased level of user control is desirable for many applications of regionalization, as researchers can ensure that the identified regions are suitable for the task at hand and do not violate any necessary constraints. For example, clusters extracted from regionalization methods may be used to define zones designated for different aspects of urban development, and it may be preferred that these zones do not cross significant geographical or infrastructural boundaries. In other applications of regionalization, however, such as identifying characteristic scales over which segregation or other socioeconomic phenomena persist \cite{wright2014patterns,olteanu2019segregation,grainger2004role,kirkley2020information}, one may be interested in imposing as few assumptions as possible about how the data clusters into regions, and instead rely on the data itself to naturally define these clusters. The minimum description length (MDL) principle from information theory is a rigorous statistical framework within which one can perform inference tasks with minimal user input \cite{grunwald2007minimum,cover2012elements}, and so provides a natural foundation for new data-driven regionalization methods.

The minimum description length principle has been applied to clustering categorical data \cite{li2004entropy}, real-valued vector data \cite{georgieva2011cluster}, and other sets of objects \cite{Kirkley22Reps} in aspatial contexts. In \cite{Rosvall07}, an algorithm for community detection in (aspatial) network data is proposed that identifies the partition minimizing the description length of an encoding of the network. This method, however, takes only topological information into account, which is relatively uninformative for planar networks of adjacent spatial regions (as is the case in regionalization). In \cite{chodrow2017structure}, a regionalization algorithm is proposed which uses concepts from information theory to define homogeneous aggregations of spatial units, which can be identified using a greedy optimization procedure. This method works well for identifying boundaries of ethnoracial segregation, but requires the user to specify the desired number of regions and chooses the class of Bregman divergences to measure information rather than a purely combinatorial description length approach.

In this paper we present a regionalization objective function for spatial networks with distributional metadata that is based solely on fundamental combinatorial arguments and the minimum description length principle. By viewing the problem of regionalization from this perspective, our approach does not require the specification of any free parameters such as an explicit dissimilarity function between spatial units or a particular value for the number of regions we want the algorithm to return. Our method also takes into account the full distribution of the covariate of interest in each spatial unit, rather than summarizing each local distribution with a single statistic such as its mode, and accounts for both this spatial metadata and the topology of regional adjacencies. We describe a greedy optimization procedure used to obtain a partition of the network that approximately minimizes this description length, which involves iteratively merging the pair of nodes that maximally reduces the description length. We then demonstrate our method on a series of experiments using both real and synthetic spatial data. In the first experiment, we illustrate how our method can effectively recover synthetically planted clusters in spatial distributional data, even in the presence of substantial noise. We move on to show that our method extracts meaningful regions and their evolution in real ethnoracial data by analyzing the New Haven-Milford metropolitan area of the U.S. as a case study, covering the decades between 1980 and 2010. Finally, in an experiment using a set of 110 large metropolitan areas across the U.S., we demonstrate that our method reveals the increasing complexity of urban segregation patterns over this same time period, and that this trend can be well explained by the increase in small scale ethnoracial diversity within these metros rather than by changes in segregation patterns at large spatial scales.


\section{Methods}
\label{sec:methods}

\subsection{Description length formulation}
\label{sec:DL}

We represent our spatial data to be regionalized as a network $G=(V,E)$ consisting of a set of spatial units (nodes) $V$ and a set of edges $E$ that connect adjacent units. More precisely, the edge $(u,v)\in E$ if and only if units $u\in V$ and $v\in V$ share a length of common border. We denote the number of units in any subset $V'\subseteq V$ of the network as $n(V')$. Over this set of $n(V)$ units, there are $b(V)\geq n(V)$ individuals residing (we adopt analogous notation for $b(V')$), and each of these individuals is classified under one of $R$ categories $r=1,2,...,R$. For example, the spatial units $u$ that comprise the network may be census tracts or block groups, and the categories could represent race, income bracket, or occupation type. We also denote with $b_r(V')$ the number of individuals of type $r$ in subset $V'\subseteq V$, such that $\sum_{r=1}^{R}b_r(V')=b(V')$. 

\begin{figure}
    \centering
    \includegraphics[width=\columnwidth]{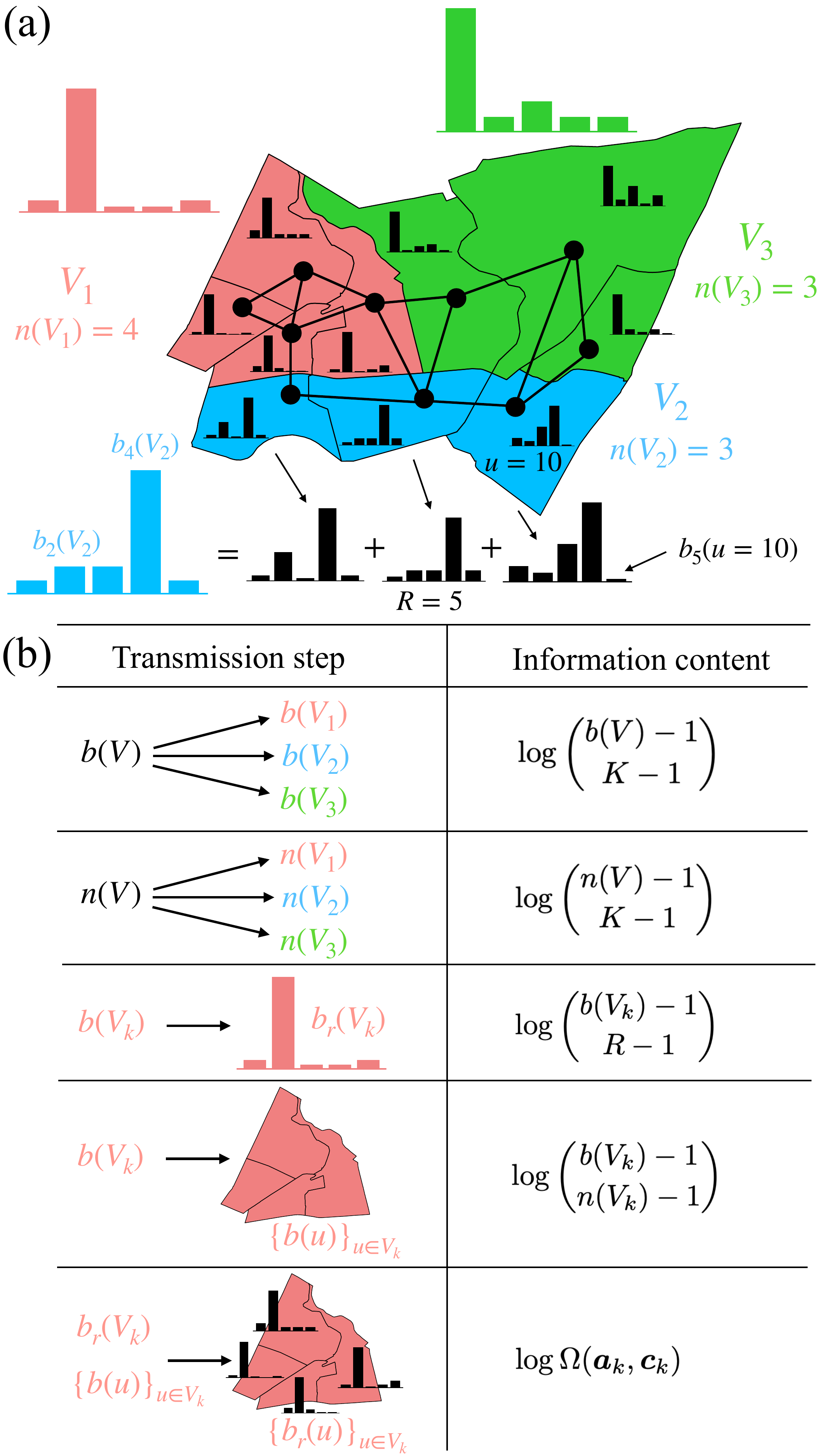}
    \caption{Diagram of description length formulation. (a) Variables used in the decription length objective (Eq.~\ref{eq:DL}), for the partition $\mathcal{P}$ of example unit-level distributions that gives the minimum description length according to Eq.~\ref{eq:DL}. The optimal contiguous partition of the underlying network of spatial adjacencies (nodes and edges in black) results in aggregated regions that capture most of the information content of the data. (b) Individual transmission steps corresponding to each of the five terms in Eq.~\ref{eq:DL}, along with their corresponding information content. Arrows go from coarser objects to more detailed subsets of these objects, which requires the specification of an amount of information quantified by the term to the right of the dividing line.}
    \label{fig:diagram}
\end{figure}

Now, suppose we want to transmit to a receiver the entire dataset $D=\{b_r(u):r=1,..,R;~u\in V\}$ consisting of the distribution of types $r$ among individuals in all units (nodes) $u\in V$. (Since we generally do not know the value $r$ for each individual due to confidentiality concerns, these unit-level distributions are the highest granularity we consider.) We will transmit this data in multiple parts, first partitioning the units $u$ into $K$ disjoint, spatially contiguous clusters $\mathcal{P}=\{V_1,V_2,...,V_K\}$ that allow us to describe the data to the receiver at a coarse spatial scale. We then transmit the small-scale details within each of these clusters by describing how the cluster's population attributes are distributed among its individual constituent units. Our goal will be to identify a partition $\mathcal{P}$ of the units such that most of the information we need to transmit is contained in the first part, or in other words, that the clusters describe most of the variation in the data and are internally homogeneous. Using the adjacency network representation $G=(V,E)$, we can guarantee spatial contiguity of the clusters by coarse-graining the network into super-nodes representing the clusters $\{V_k\}$ through merging nodes in $V$ that share edges in $E$. A diagram of a partition $\mathcal{P}$ of an example network and a list of the variables used in the information transmission scheme are shown in Fig.~\ref{fig:diagram}a.  

We assume that the receiver knows there are $n(V)$ units in total that will be assigned to $K$ clusters, and that there are $b(V)$ individuals with $R$ distinct categories that will be assigned to units $u\in V$. (Transmitting these requires a negligible amount of information, so we can safely ignore them in our description length anyway.) We first need to transmit the populations $b(V_k)$ for each of the clusters $V_k$, which consists of a configuration of $K$ non-negative integer values that sum to $b(V)$. Prior to transmission of the data $D$, we must develop a common codebook with the receiver, from which we will transmit a binary string representing the particular configuration of the populations $\{b(V_k)\}$. Assuming $K\ll b(V)$, there are approximately ${b(V)-1 \choose K-1}$ possible configurations of these values we must encode, and so we will possibly have to send a bitstring of length $\lceil\log_2 {b(V)-1 \choose K-1}\rceil$ to the receiver to transmit the cluster-level populations $\{b(V_k)\}$. ($\lceil x\rceil$ denotes the smallest integer not less than $x$, and we will omit this transformation in future considerations as its contribution is negligible for $x\gg 1$. For the sake of brevity we will also denote $\log_2(x)\equiv \log(x)$.) Thus, the information content (or ``description length'') of this step in the transmission procedure is 
\begin{equation}
\label{eq:Lbvk}
\mathcal{L}(\{b(V_k)\}) = \log {b(V)-1 \choose K-1}. \end{equation}

Following the same logic, we can construct the description lengths for the rest of the steps required to transmit $D$ according to this scheme. After sending the populations $\{b(V_k)\}$, we must transmit the number of units within each cluster, $\{n(V_k)\}$, for which we will construct a different codebook. This step will have a description length of the same form as Eq.~\ref{eq:Lbvk}, thus 
\begin{equation}
\label{eq:Lnvk}
\mathcal{L}(\{n(V_k)\}) = \log {n(V)-1 \choose K-1}.   
\end{equation}
Now, for each cluster $V_k$ we need to transmit the size distribution $\{b_r(V_k)\}$ of categories within the population $b(V_k)$, which will have the same form as Eq.s~\ref{eq:Lbvk} and \ref{eq:Lnvk}. The description length of this step will be a sum over such description lengths, or
\begin{equation}
\label{eq:Lnrvk}
\mathcal{L}(\{b_r(V_k)\}) = \sum_{k=1}^{K}\log {b(V_k)-1 \choose R-1}.   
\end{equation}
Similarly, we need to transmit the populations $b(u)$ of the units $u\in V_k$, for each cluster $V_k$, which will give a total description length contribution of
\begin{equation}
\label{eq:Lbu}
\mathcal{L}(\{b(u)\}) = \sum_{k=1}^{K}\log {b(V_k)-1 \choose n(V_k)-1}.   
\end{equation}

The receiver now knows how many units $u$ are in each cluster $V_k$, how many individuals are in each of these units, and how categories are distributed across the entire population of each $V_k$. The only information left to transmit is how the categories in each cluster $V_k$ are distributed among the populations in $V_k$'s constituent units $u$. (We ignore the information required to map the final unit-level distributions to particular locations in the network.) The number of ways these values can be distributed is equivalent to the number $\Omega(\vec{a}_k,\vec{c}_k)$ of non-negative integer-valued matrices with row sums $\vec{a}_k=\{b(u)\}_{u\in V_k}$ and column sums $\vec{c}_k=\{b_r(V_k)\}_{r=1}^{R}$. We can see this by noting that there are $b(V_k)$ total individuals in cluster $V_k$, and using the identities
\begin{equation}
b(V_k) = \sum_{u\in V_k}b(u)    
\end{equation}
and
\begin{equation}
b(V_k) = \sum_{r=1}^{R}b_r(V_k).    
\end{equation}
The description length for this final step is thus given by
\begin{equation}
\label{eq:Lfinal}
\mathcal{L}_{final} = \sum_{k=1}^{K}\log \Omega(\vec{a}_k,\vec{c}_k).    
\end{equation}
Computing $\Omega(\vec{a}_k,\vec{c}_k)$ is in general challenging, but it can be approximated in the regime $R,n(V_k)\ll b(V_k)$, which is typically the regime we encounter in practice (see Ref.~\cite{NCY20} for details on this approximation). 

Taken all together, the total description length of the data $D$ under the partition $\mathcal{P}$ of the network $G$ is given by the sum of Eq.s~\ref{eq:Lbvk}, \ref{eq:Lnvk}, \ref{eq:Lnrvk}, \ref{eq:Lbu}, and \ref{eq:Lfinal}, thus
\begin{equation}
\label{eq:DL}
\begin{split}
\mathcal{L}(D,\mathcal{P}) &= \log {b(V)-1 \choose K-1} + \log {n(V)-1 \choose K-1}\\
&~~+ \sum_{k=1}^{K}\log {b(V_k)-1 \choose R-1} + \sum_{k=1}^{K}\log {b(V_k)-1 \choose n(V_k)-1} \\
&~~+ \sum_{k=1}^{K}\log \Omega(\vec{a}_k,\vec{c}_k).
\end{split}
\end{equation}
A list of the individual transmission steps and their corresponding information content contribution to Eq.~\ref{eq:DL} is shown in Fig.~\ref{fig:diagram}b.

We can see that the first three terms in Eq.~\ref{eq:DL} penalize us for having a greater number of clusters $K$, as they will tend to contribute greater description lengths as $K$ increases, and the fourth term will not depend on the number of clusters to first order in a Stirling approximation of the binomial coefficients. For the last term in Eq.~\ref{eq:DL}, in the extreme case where there is only one category $r^\ast$ that is represented in the population of the units $u\in V_k$ (i.e. $\vec{c}_k[r]=0$ for $r\neq r^\ast$), then we have $\Omega(\vec{a}_k,\vec{c}_k)=1$ and the contribution from this term vanishes. More generally, there are fewer ways the categories can be distributed among the populations in $V_k$'s constituent tracts if $\vec{c}_k$ is more concentrated on a single category, and so the last term in Eq.~\ref{eq:DL} will penalize us for having a high level of diversity within the clusters. (Or, conversely, this term encourages partitions $\mathcal{P}$ that have homogeneous clusters.)    

The optimal partition $\mathcal{P}=\{V_1,...,V_k\}$ of the network $G$ that minimizes the description length in Eq.~\ref{eq:DL} will allow us to communicate most of the information about the data $D$ through the cluster-level distributions alone, but penalize us for constructing these clusters at too small a scale, since this will not save us much effort above and beyond simply transmitting all the unit-level data individually. The goal of our regionalization algorithm is to identify this partition, and we describe an algorithm to accomplish this task in the next section.

\subsection{Optimization and model selection}
\label{sec:optimization}

Minimization of the description length in Eq.~\ref{eq:DL}, like many other regionalization objectives \cite{duque2007supervised}, is a combinatorial optimization problem that can be approached in a number of ways to obtain an approximate solution. Here, we opt for a greedy solution that consists of starting with each node in its own cluster then iteratively merging the pair of adjacent clusters whose aggregation results in the largest decrease in Eq.~\ref{eq:DL}, until no merges produce a negative change in the description length. We consider the two clusters of units $V_k$ and $V_{k'}$ adjacent if and only if there exists a $u\in V_k$ and $v\in V_{k'}$ such that $(u,v)\in E$. This merging procedure thus has the benefit of naturally ensuring that the partition $\mathcal{P}$ produces only contiguous clusters of units, since if units $u$ and $v$ end up in the same cluster $V_k$, there must be a path of edges in $E$ that connect $u$ and $v$ such that all nodes along this path are also in $V_k$. 

For any pair of clusters $V_k$ and $V_{k'}$, we can quickly compute the change in Eq.~\ref{eq:DL} that results from their aggregation into a single cluster, $V_{k,k'}$. Supposing there are $K$ clusters prior to the proposed merge, the change in description length from merging $V_k$ and $V_{k'}$ is given by 
\begin{equation}
\label{eq:deltaDL}
\begin{split}
\Delta \mathcal{L}(k,k') &= \log {b(V_{k,k'})-1 \choose R-1}+ \log {b(V_{k,k'})-1 \choose n(V_{k,k'})-1}   \\
& + \log \Omega(\vec{a}_{k,k'},\vec{c}_{k,k'}) - \log {b(V_{k})-1 \choose R-1} \\
&- \log {b(V_{k})-1 \choose n(V_{k})-1} - \log \Omega(\vec{a}_{k},\vec{c}_{k})  \\
&- \log {b(V_{k'})-1 \choose R-1} - \log {b(V_{k'})-1 \choose n(V_{k'})-1}   \\
&- \log \Omega(\vec{a}_{k'},\vec{c}_{k'}).
\end{split}
\end{equation}
Here we have ignored the first two terms in Eq.~\ref{eq:DL}, as these terms change by the same amount across all pairs $k,k'$ and thus do not need to be computed until the optimal pair $k,k'$ is chosen. (Whether or not this pair will be merged or the algorithm will terminate does depend on these first two terms, which can be computed in constant time.) This expression can be evaluated in $\Ord(n(V_k)+n(V_{k'}))$ time for each pair of clusters $k,k'$. Additionally, it only needs to be computed once for each pair, and can be reused for future iterations of the algorithm if the pair $k,k'$ does not get merged (as long as each newly formed cluster gets a unique label). Once no remaining pair of clusters can be merged to reduce the description length ($\Delta \mathcal{L}(k,k')>0$ for all adjacent pairs $V_k,V_{k'}$), the algorithm terminates.

The adjacency relations between clusters are updated as the algorithm progress by considering the clusters as ``super-nodes'' whose neighbor sets are merged at each step. This takes an additional $\Ord(d_k+d_{k'})$ operations, where $d_k$ is the number of adjacent clusters (super-nodes) to cluster (super-node) $k$, and is typically smaller than $\Ord(n(V_k)+n(V_{k'}))$ for large clusters, since many clusters are adjacent to only a few others for planar graphs (this is not necessarily the case for non-planar networks). We find in practice that the algorithm scales well to large systems, running in less than order $\Ord(n(V)^2)$ time for the entire clustering procedure (see Appendix~\ref{sec:runtime} and Fig.~\ref{fig:runtime}). 

Although the greedy algorithm used to optimize the description length in Eq.~\ref{eq:DL} has the advantages of being computationally efficient and simple to implement, it is not guaranteed to identify the true optimal partition $\mathcal{P}$ that minimizes the description length objective over all possible partitions of the network into contiguous regions. Identifying the optimal partition $\mathcal{P}$ is a computationally challenging optimization problem, as there are at least $\Ord(n(V)^2)$ (and at worst exponentially many) contiguous partitions of the network one must account for \cite{vince2017counting}, and even sampling such partitions is itself intractable for planar graphs \cite{najt2019complexity}. Additionally, fast dynamic programming approaches used for exactly solving contiguous clustering problems in one dimension are not applicable \cite{wang2011ckmeans}. However, we find in test examples that the greedy algorithm gives results quite competitive with those obtained through exhaustive enumeration of all contiguous partitions of the network to identify the true optimal partition (see Appendix~\ref{sec:exactcomparison} and Fig.~\ref{fig:exactcomparison}). 

The first few terms in Eq.~\ref{eq:DL} penalize us for having a large number of clusters, since we waste information describing all of the cluster-level distributions in their entirety. Meanwhile, the last term penalizes us for having a small number of clusters, since we waste information describing the small scale details of these clusters when they encompass too broad a variety of unit-level distributions. The optimal balance, and thus the optimal value of $K$, lies somewhere in between with an intermediate number of clusters, and the description length in Eq.~\ref{eq:DL} thus performs model selection for $K$ automatically. In our example applications, we therefore choose to let the description length tell us exactly how many clusters are in the data. However, in many applications it may be preferable to have a fixed value of $K$ \cite{duque2007supervised}, and this can easily be accommodated in our algorithm by simply performing the greedy merge moves until the desired number of clusters is reached. 

We can assess the quality of the information compression achieved through partitioning the units into clusters by comparing the final description length $\mathcal{L}(D,\mathcal{P})$ for the optimal partition $\mathcal{P}$ with the description length $\mathcal{L}(D,\mathcal{P}_0)$ for the trivial partition $\mathcal{P}_0$ in which each unit is in its own cluster (computed at the beginning of the optimization algorithm). From this we can construct an inverse ``compression ratio'' for the data $D$ as
\begin{equation}
\label{eq:comp}
\eta(D) = \frac{\text{compressed size of $D$}}{\text{uncompressed size of $D$}} =\frac{\mathcal{L}(D,\mathcal{P})}{\mathcal{L}(D,\mathcal{P}_0)}.    
\end{equation}
$\eta(D)$ approaches its minimum value of $0$ when the data $D$ can be compressed extremely efficiently through partitioning the network $G$, and approaches its maximum value of $1$ when there is no partition of $G$ that achieves any compression of $D$. 

Eq.~\ref{eq:comp} can thus be used as a measure of the complexity of the spatial segregation of the data $D$, with more complex spatial distributions of the covariate of interest resulting in higher inverse compression ratios $\eta$. Intuitively, if the data $D$ is very easy to compress (low $\eta$), then it is highly spatially segregated into homogeneous clusters, and most of the information in $D$ is captured at large scales. On the other hand, if the data is very hard to compress (high $\eta$), then much of the information in the data is manifested at small spatial scales, which could be due to the presence of diversity at these small spatial scales among other factors that contribute to the multifaceted spatial nature of segregation patterns \cite{massey1988dimensions}. The inverse compression ratio in Eq.~\ref{eq:comp} also allows us to compare the compressibility of datasets with different populations $b(V)$, numbers of categories $R$, number of spatial units $n(V)$, or where categories are defined differently. Indeed, for $b(V)\gg n(V) \gg R,K$---which we typically encounter in practice for demographic data---the leading order scaling of both $\mathcal{L}(D,\mathcal{P})$ and $\mathcal{L}(D,\mathcal{P}_0)$ in Eq.~\ref{eq:comp} is $\Ord(nR\log b)$.


\subsection{Ethnoracial data in U.S. metropolitan areas}
\label{sec:data}

To examine the performance of our algorithm in a practical context we test our method using ethnoracial data that take the form of distributions within census tracts. Ethnoracial distributions for census tracts in U.S. metro areas were obtained from the Longitudinal Tract Database \cite{logan2014interpolating}, which maps 2010 census tract boundaries to ethnoracial distribution data for decades going back to 1970. (Data from 1970 are omitted from our analysis, as they do not include the designation of Hispanic ethnicity.) The race/ethnicity categories considered are `Non-Hispanic White', `Non-Hispanic Black', `Asian', `Hispanic', and `Other', which includes persons not categorized under the first four groups.

To process the census tract networks for each metropolitan area, we first map each census tract to its corresponding core-based Metropolitan Statistical Area (MSA) using the county designation of the tract. MSA's are used as the metro regions for this analysis as they aim to encompass areas of unified social and economic labor market forces, while also enclosing full counties which allows us to avoid splitting census tracts \cite{bettencourt21}. It is important to be mindful of this choice of metro regions, since the Modifiable Areal Unit Problem can result in different conclusions about city-level socioeconomic diversity depending on which boundaries are chosen \cite{gehlke1934certain,cottineau2017diverse}. 

We then use TIGER shapefile data \cite{united2019tiger} for the census tracts to determine the network $G=(V,E)$ of adjacent tracts in each MSA. Finally, the longitudinal ethnoracial distribution data is then mapped to the nodes in each network using the census tract IDs. To reduce noise as much as possible in our analysis, we kept only metros with at least 100 tracts that had complete ethnoracial distribution estimates in all tracts for the four decades 1980, 1990, 2000, and 2010. After preprocessing, 110 metro networks remained for the analysis in Sec.~\ref{sec:metros}, one of which was the New Haven-Milford metro used for the case study in Sec.~\ref{sec:casestudy}. We make the tract adjacency networks for each metro we used in our analysis (with accompanying node metadata including ethnoracial distributions), as well as code for executing our algorithm publicly available at \url{https://github.com/aleckirkley/MDL_regionalization}.


\section{Results}
\label{sec:results}

\subsection{Cluster recovery in synthetic data}
\label{sec:synthetic}

As a first test of our method, we explore its capability of recovering clusters in synthetic data. To do this, we create a synthetic model of spatial distributional data that has four tunable parameters: the number of clusters $K$, the number of covariate categories $R$, the level of statistical noise between the cluster-level distributions $\epsilon_{between}$, and the level of statistical noise within the clusters, $\epsilon_{within}$. The model requires a spatial network $G=(V,E)$ representing the adjacencies among spatial units, and for this we use the census tract network for the New Haven-Milford metropolitan area, with $n(V)=189$ census tracts (see Sec.~\ref{sec:data} for details). (The specific choice of $G$ does not tend to make a qualitative difference in the results, since the spatial networks induced by the adjacencies between units will in general have very restricted topologies \cite{barthelemy2011spatial}.) It is also possible to include variable unit populations $b(u)$ in this model, but for simplicity we set $b(u)=10000$ for all $u\in V$ so that these values correspond roughly to the values seen in the real U.S. census tract data used in Sec.~\ref{sec:casestudy} and Sec.~\ref{sec:metros}. We show that this population heterogeneity has little effect on downstream results in Appendix~\ref{sec:populationhet} and Fig.s \ref{fig:equalcomparisonscale} and \ref{fig:equalcomparisonavgH}.

To generate a realization of the model, we first randomly partition the units into contiguous clusters by picking $K$ units (``seeds'') at random and constructing the Voronoi tesselation of the centroids of the spatial units of the network with respect to these seeds. This Voronoi tesselation places each unit into the cluster corresponding to the seed geographically nearest to the unit's centroid in terms of Euclidean distance, and in doing so tends to produce clusters are spatially contiguous (we reject the proposed partition if it has any discontiguous partitions). The Voronoi tesselation produces relatively compact convex regions in the plane, but there are other reasonable alternative tesselations for generating the randomized contiguous partition. We denote this ``planted'' partition $\mathcal{P}_{planted}$, to distinguish it from the partition $\mathcal{P}$ inferred using our minimum description length algorithm. 

Next, each cluster $V_k$ is assigned a vector $\vec{x}(V_k)$ which tunes the covariate distributions within the units that comprise $V_k$. $\vec{x}(V_k)$ is drawn from a Dirichlet distribution with length-$R$ concentration parameter $\vec{\alpha}=\epsilon_{between}^{-1}\vec{1}_{R}$. This allows us to tune the level of differentiation between the cluster-level distributions, as well as the localization of these distributions. For low levels of between-cluster noise $\epsilon_{between}$ ($\epsilon_{between}\lesssim 1$), the distributions $\vec{x}(V_k)$ will all tend to distribute their probability relatively equally around the $R$ categories, and there is little differentiation between the clusters $V_k$. On the other hand, for high levels of between-cluster noise $\epsilon_{between}$ ($\epsilon_{between}\gtrsim 5$), there will be high between-cluster variance in the distributions $\{\vec{x}(V_k)\}$, which will each tend to localize around a single category $r$. In general, the higher the between-cluster noise $\epsilon_{between}$ is, the easier it should be to recover the planted clusters in the synthetic data with our partitioning algorithm, since the clusters are more easily distinguished. 

To tune the level of noise within each cluster $V_k$, we generate the distribution $\vec{x}(u)=\{b_r(u)\}_{r=1}^{R}/b(u)$ for each $u\in V_k$ using $\vec{x}(u)=(1-\epsilon_{within})\vec{x}(V_k)+\epsilon_{within}\vec{x}_{noise}$, where $\vec{x}_{noise}$ is drawn from a Dirichlet distribution with concentration parameters equal to $1$. If the level of within-cluster noise $\epsilon_{within}\approx 0$, then each $\vec{x}(u)$ for $u\in V_k$ will be roughly the same as $\vec{x}(V_k)$, and thus the unit-level distributions $\{b_r(u)\}$ for $u\in V_k$ are very similar. On the other hand, if the level of within-cluster noise $\epsilon_{within}\approx 1$, then the vectors $\vec{x}(u)$ will have high variability within the cluster $V_k$ and the distributions $\{b_r(u)\}$ for $u\in V_k$ will share very little information. As opposed to the between-cluster noise, higher values of the within-cluster noise $\epsilon_{within}$ correspond to it being \emph{harder} to recover the planted clusters in the synthetic data, since the unit-level distributions within each clusters are not as similar to each other. Illustrative examples of realizations of this synthetic data model used for the experiments in this section are shown in Fig.~\ref{fig:syntheticdiagram}.

To measure the performance of our algorithm for any particular draw from the model, we compute the normalized mutual information \cite{vinh2010information} between our inferred minimum description length partition $\mathcal{P}$ and the planted partition $\mathcal{P}_{planted}$. The mutual information tells us how much information is shared between the two partitions, and its value is then normalized to fall in $[0,1]$ so that $0$ corresponds to completely uncorrelated partitions, and $1$ corresponds to identical partitions (up to an arbitrary relabeling of the clusters). Letting $\mathcal{P}=\{V_{k}\}$ and $\mathcal{P}_{planted}=\{U_{k'}\}$,
the mutual information $MI(\mathcal{P},\mathcal{P}_{planted})$ is given by
\begin{equation}
MI(\mathcal{P},\mathcal{P}_{planted}) = \sum_{k,k'}\frac{\abs{V_k\cap U_{k'}}}{n(V)}\log\frac{n(V)\abs{V_k\cap U_{k'}}}{\abs{V_k}\abs{U_{k'}}}.    
\end{equation}
The mutual information can be normalized to fall in $[0,1]$ by dividing by the average of the entropies of the individual partitions $\mathcal{P}$ and $\mathcal{P}_{planted}$, giving
\begin{equation}
\label{eq:NMI}
NMI(\mathcal{P},\mathcal{P}_{planted}) = 2\frac{MI(\mathcal{P},\mathcal{P}_{planted})}{H(\mathcal{P})+H(\mathcal{P}_{planted})},
\end{equation}
with
\begin{equation}
H(\mathcal{P}) = -\sum_{k}\frac{\abs{V_k}}{n(V)}\log \frac{\abs{V_k}}{n(V)}   
\end{equation}
and
\begin{equation}
H(\mathcal{P}_{planted}) = -\sum_{k'}\frac{\abs{U_{k'}}}{n(V)}\log \frac{\abs{U_{k'}}}{n(V)}.    
\end{equation}

The normalized mutual information is a standard and well-tested measure for comparing partitions of networks \cite{DDDA05,lancichinetti2008benchmark}, but it has a critical shortcoming for our particular application in that it gives very high baseline values to completely random contiguous partitions of spatial networks. The reason for this is that Eq.~\ref{eq:NMI} compares the partitions $\mathcal{P}$ and $\mathcal{P}_{planted}$ relative to the ensemble of all possible partitions of the network, contiguous or not, and the constraint of contiguity induces a high baseline level of correlation between the partitions. To correct for this, we rescale the normalized mutual information by subtracting off its maximum value at $\epsilon_{within}=1$ over all simulations, which we denote $NMI_{baseline}$, and dividing by one minus this baseline value. The resulting measure is more appropriate for comparing spatially contiguous partitions, and is given by
\begin{equation}
\label{eq:NMIrescaled}
NMI_{rescaled} = \frac{NMI - NMI_{baseline}}{1-NMI_{baseline}}.
\end{equation}
It is then easy to see when we reach the NMI value at which the partitions are minimally correlated, subject to the contiguity constraint, since the rescaled measure in Eq.~\ref{eq:NMIrescaled} will be near $0$. Our rescaling does not map the highest value of the NMI over the $\epsilon_{within}$ range in a given experiment to $1$, so that we have better differentiation of performance in the low noise region. Indeed, we will see that the zero-noise values of the rescaled NMI are slightly less than $1$ in most cases, since some sampled model realizations will by chance produce some adjacent clusters that are nearly indistingushable.

\begin{figure}
    \centering
    \includegraphics[width=\columnwidth]{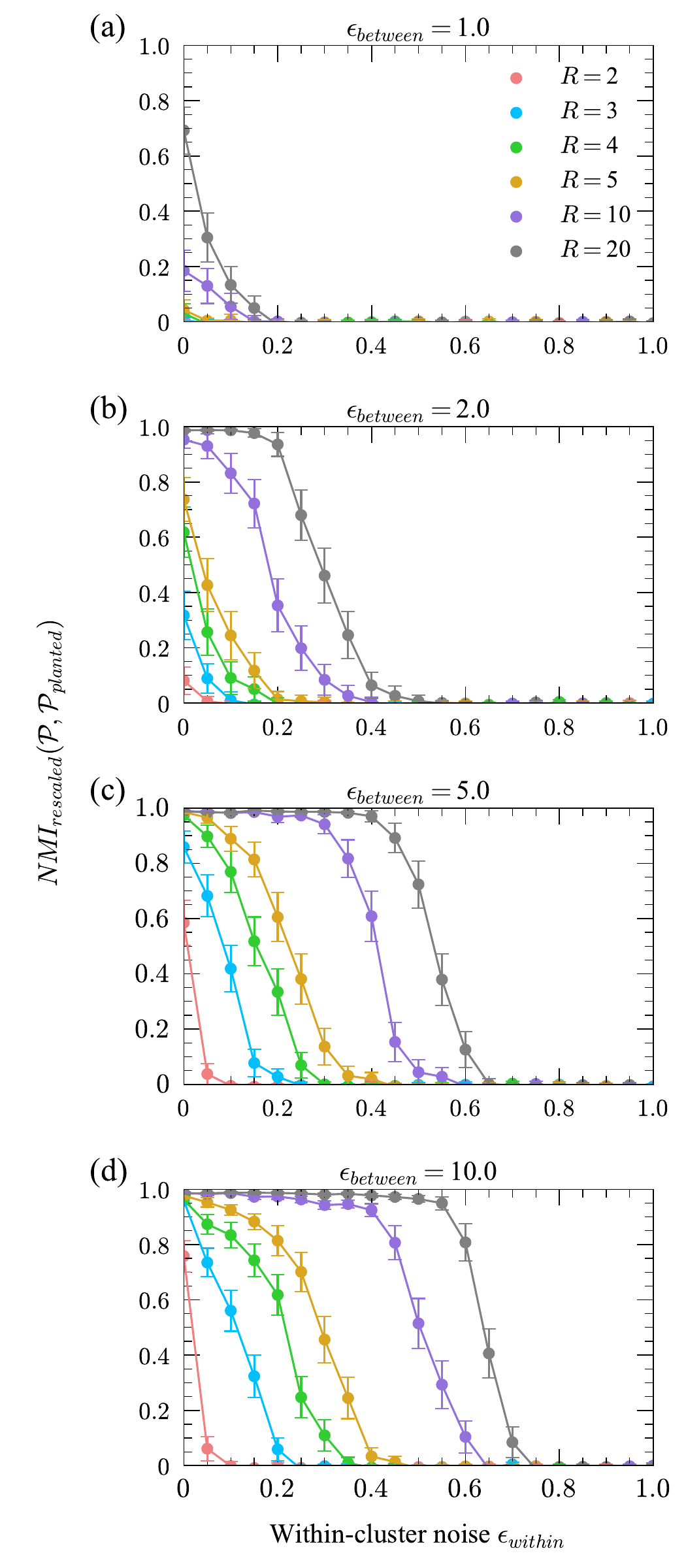}
    \caption{Recovery of synthetic clusters generated by the model in Sec.~\ref{sec:synthetic}. In each panel, the recovery performance of our algorithm, as measured by the rescaled NMI of Eq.~\ref{eq:NMIrescaled}, is plotted against the level of within-cluster noise $\epsilon_{within}$, for various values of between-cluster noise $\epsilon_{between}$. The number of covariate categories $R$ is varied across the panels, and the number of clusters is set to $K=5$.} 
    \label{fig:synthetic}
\end{figure}

In Fig.~\ref{fig:synthetic} we show the results of generating realizations of synthetic contiguous partitions from our model and running our regionalization algorithm on each of these realizations to try to recover the planted clusters. To summarize the distribution of results over the ensemble of planted partitions generated from the model, each data point represents the average rescaled normalized mutual information over $100$ of these cluster recovery experiments, with error bars representing $2$ standard errors in the mean. We can see that as the level of within-cluster noise $\epsilon_{within}$ increases, it becomes harder for us to recover the planted partition (as expected), but that we still have recovery better than the baseline value for reasonably high levels of within-cluster noise, for $\epsilon_{between} > 1$. (At $\epsilon_{between}=1$, there is not enough differentiation in the latent cluster-level distributions $\bm{x}(V_k)$ for a distinguishable cluster structure except for at very low levels of within-cluster noise $\epsilon_{within}$.) As expected, we can observe that the recovery task becomes easier as $\epsilon_{between}$ increases, since we have better differentiation in the latent cluster-level distributions $\{\bm{x}(V_k)\}$. We can see that the exact values of $\epsilon_{within}$ and $\epsilon_{between}$ at which significant enough noise is introduced to obscure the cluster structure of the data are different, since $\epsilon_{within}\in [0,1]$ is a fractional weight and $\epsilon_{between}\in [0,\infty)$ is an inverse Dirichlet concentration parameter. Recovery performance also improves as $R$ increases, as it is less likely for the modes of the distributions $\vec{x}(V_k)$ to overlap for larger $R$. The performance of our algorithm does not vary significantly with the number of planted clusters $K$, so results are displayed only for $K=5$ for clearer visualization.

Overall, the results of Figure~\ref{fig:synthetic} indicate that our minimum description length regionalization algorithm is able to successfully recover artificially planted clusters, even in the presence of substantial noise, with the performance varying as expected with the level of homogeneity within and between clusters. We now move on to examine its performance on real ethnoracial distribution data.


\subsection{Case study: Ethnoracial composition of the New Haven-Milford metropolitan area}
\label{sec:casestudy}

\begin{figure}
    \centering
    \includegraphics[width=\columnwidth]{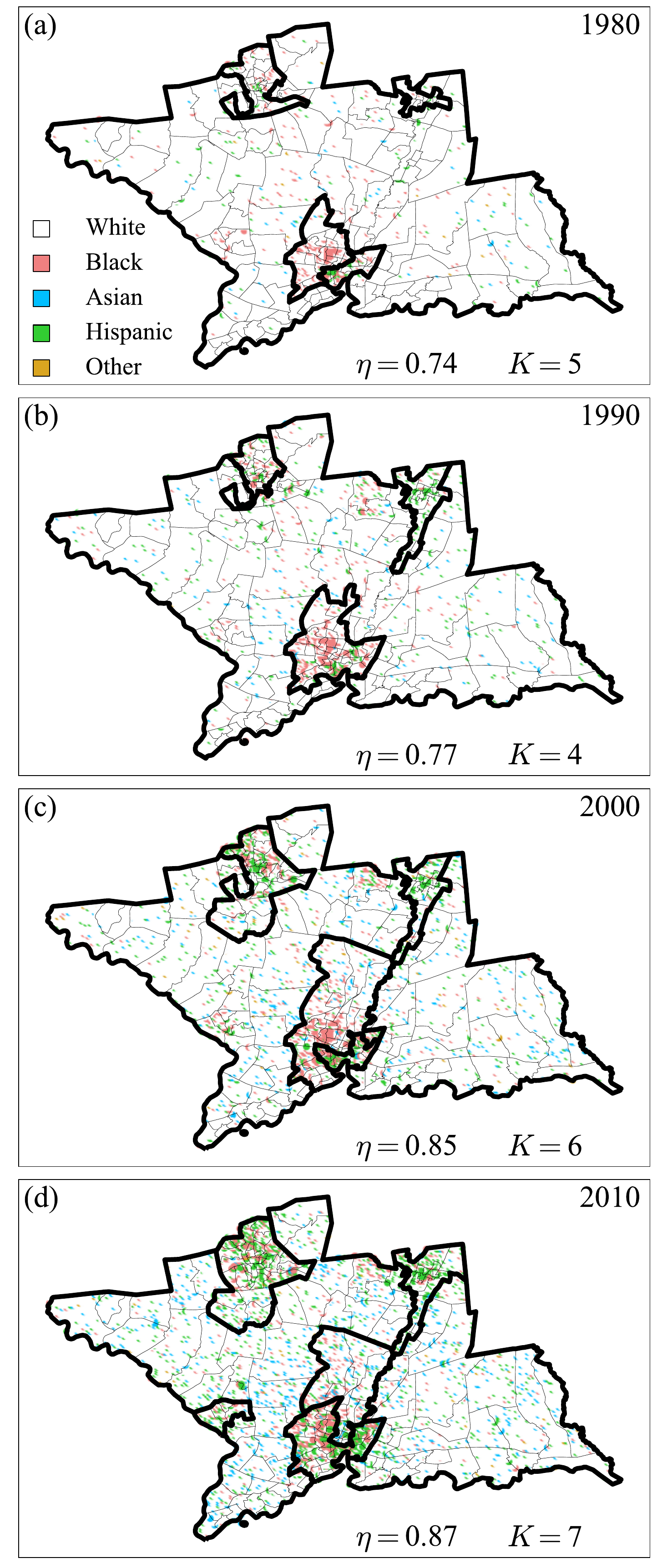}
    \caption{Ethnoracial distributions in census tracts (thin black borders) within the New Haven-Milford, Connecticut metropolitan area, with inferred cluster boundaries from the minimum description length regionalization algorithm (thick black borders). Colored points are distributed at random within each tract and each color covers an area proportional to the fraction of the population within the tract that falls under the ethnoracial category corresponding to that color. The inverse compression ratio $\eta$ (Eq.~\ref{eq:comp}) and the optimal number of clusters $K$ are shown for each decade.}
    \label{fig:NHclusts}
\end{figure}

To illustrate how the clusters obtained with our regionalization algorithm capture meaningful patterns in real data, we look at a case study of the ethnoracial evolution of the New Haven-Milford, Connecticut metropolitan area, using the data described in Sec.~\ref{sec:data}. This metro was chosen for the case study analysis due to a clearly visible spatial evolution of different ethnoracial groups and relatively low heterogeneity in census tract density in comparison with other smaller metros in our dataset, both factors allowing for a clear visual analysis of its temporal segregation patterns. Additionally, the New Haven-Milford metro exhibits a noticeable increase in ethnoracial diversity at small scales, which will help us motivate the analysis in Sec.~\ref{sec:metros}.   

In Fig.~\ref{fig:NHclusts}, we show the evolution of the spatial distribution of ethnoracial groups, along with the regional boundaries inferred from minimizing the description length in Eq.~\ref{eq:DL}, for the census tracts in the New Haven-Milford metro area between 1980 and 2010. Points are distributed randomly within each tract in proportion to the fraction of the population in each ethnoracial category. We can see that, in general, the clusters inferred through our algorithm correspond to heterogeneities in the spatial densities of these ethnoracial groups. The outlying tracts in the clusters, particularly in the year 2000, do not have as high a proportion of minority ethnoracial groups as the more densely packed areas of the clusters, but we can see these areas begin to fill out with minority populations over time. (Their inclusion status in the cluster is determined by their slightly higher relative concentrations of the minority groups dominant in the core of their cluster, compared to nearby areas.)     

Two emerging Black/Hispanic clusters in the north and one in the south are the primary clusters dense with minority populations that are captured by the algorithm, which assigns the rest of the metro to a single more rural/suburban and predominantly White cluster in all years (in 2000 and 2010 this cluster is broken into two due to contiguity requirements). We see that these clusters trend towards higher percentages of Hispanics relative to Non-Hispanic Blacks, which is consistent with the high influx of Latinos to the area between 1990 and 2000 \cite{NewHaven2003}. The spatial extent of these Black/Hispanic clusters increases over time, reaching out into the less dense region of the metro that was predominantly White in 1980, which is consistent with `White flight' during deindustrialization as well as the expanding influence of Yale University in the south \cite{di2006there}. In 2010, we see a slightly different configuration of clusters, with the northern Black/Hispanic clusters remaining largely intact, but the southern-most cluster splitting into a largely Black/Hispanic cluster and one relatively mixed cluster. In 2000, this mixed cluster was merged with a primarily Black cluster, but in 2010 we can see that the movement of Hispanic population into the previously Black cluster provided a high enough level of Black/Hispanic mixing to create a single dense southern-most cluster, and a separate cluster to the north with smaller overall minority populations. In 2010 we also see the emergence of a new largely Hispanic cluster to the west. These emerging clusters are reflected by an increasing optimal number of clusters, $K$, over the last three decades. The high level of spatial aggregation of Hispanic populations we see in the New Haven-Milford metro area is consistent with a general trend revealed by a fractal scaling analysis of large U.S. cities \cite{stepinski2020complexity}, which found that from 1990 to 2010 the fractal dimensions of predominantly Hispanic areas increased in most of the cities studied. 

In addition to the emergent Black/Hispanic clusters at larger spatial scales, the rural/suburban tracts diversified metro-wide due to an influx of Asian and Hispanic populations to the area \cite{NewHaven2015}. For the most part these outlying tracts do not have sufficient differentiation in their ethnoracial distributions to necessitate separate clusters, and they are all grouped into a similar majority-White cluster for all four decades. However, this increasing tract-level diversity does result in greater difficulty compressing the data, as there is a clear positive trend in the inverse compression ratio $\eta$ (Eq.~\ref{eq:comp}) over the four decades. We will show in the next section that a similar trend is seen across all the metros in our dataset, and that this decreasing compressibility can be better attributed to the latter effect observed in this case study (small-scale diversification) than the former (changes in large scale segregation).


\subsection{Compression of ethnoracial data across metros}
\label{sec:metros}

\begin{figure}
    \centering
    \includegraphics[width=\columnwidth]{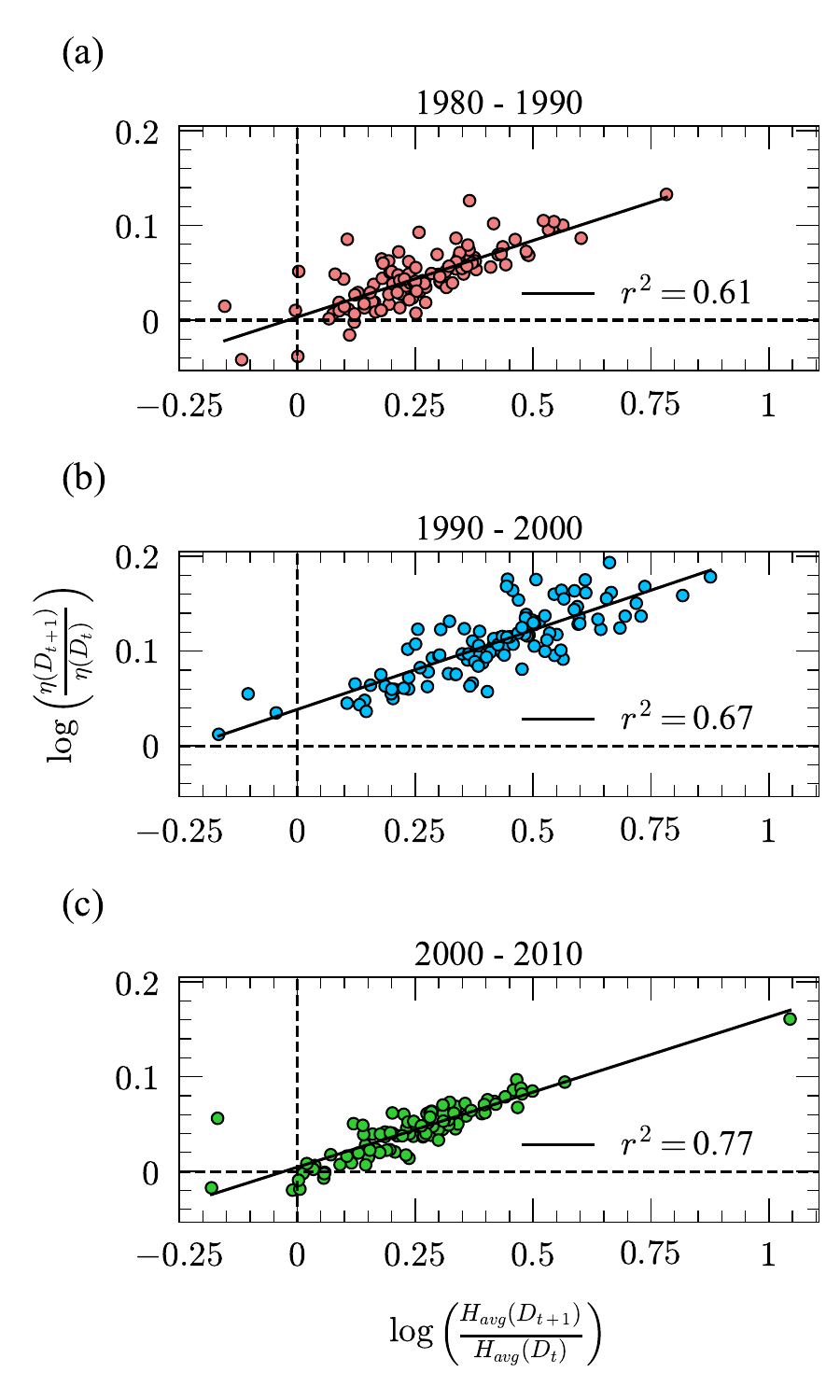}
    \caption{Log-ratio of consecutive inverse compression ratios $\eta$ (Eq.~\ref{eq:comp}) versus the log-ratio of consecutive average tract-level diversities $H_{avg}$ (Eq.~\ref{eq:Havg}), in U.S. metros over the decades spanning $t=1980$ to $t=2010$. Dotted lines at $x=0$ and $y=0$ are displayed for reference, along with OLS regression lines (solid black) and their coefficients of determination $r^2$. The slopes of all regression lines were highly statistically significant at the $0.01$ significance level. Without grouping the changes by decade, we find $r^2=0.70$, and that the slope is again highly significant at the $0.01$ significance level.}
    \label{fig:entvcomp}
\end{figure}

Now that we have demonstrated that our regionalization method is capable of identifying meaningful clusters in ethnoracial census data, we move on to a large scale analysis of the metro area networks described in Sec.~\ref{sec:data}. Specifically, we look at the extent to which the data within each metro can be compressed by our algorithm according to Eq.~\ref{eq:comp}, which as discussed in Sec.~\ref{sec:optimization} can be used as an indicator of the overall complexity of the segregation patterns in these areas. 

From a purely visual analysis, one can easily argue that the segregation patterns seen in the New Haven-Milford metro in Fig.~\ref{fig:NHclusts} are becoming more complex over time: describing to somebody the spatial distribution of ethnoracial groups in this metro would require more effort in 2010 than in 1980. And although this concept is difficult to express in precise language due to the highly multifaceted nature of patterns in spatial data, we can capture this intuition through the inverse compression ratio of Eq.~\ref{eq:comp}, which tells us how efficiently we can compress the data for its exact description to a receiver.

Despite the difficulty we may have in succinctly articulating the overall complexity of the observed segregation patterns, there are a few key features that stand out in the plots of Fig.~\ref{fig:NHclusts}. As discussed in Sec.~\ref{sec:casestudy}, the inverse compression ratio $\eta(D)$ increases for the New Haven-Milford over the decades spanning 1980 to 2010, and it is uncertain whether or not this increase can be better attributed to changes in tract-scale diversity or changes in large-scale segregation. The first feature of interest is the increasing diversity of a typical tract in the metro area, demonstrated by a greater and greater fraction of area covered by colored points as time progresses. The second feature that stands out is the changing spatial extent of the clustered areas, seen through the gradual absorption of the primarily White outlying tracts in 1980 into the minority-dense clusters as these clusters expand. In this section we explore the question of whether or not spatial ethnoracial patterns become more complex (as quantified by Eq.~\ref{eq:comp}) in metros other than New Haven-Milford, and to what extent the patterns we observe across these metros are consistent with each of these two features of overall diversity and changing spatial scales of clustering. 

To measure the tract-level diversity of the data $D$ in each metro area, the first feature of interest, we compute the average entropy $H_{avg}(D)$ of the ethnoracial distribution in each tract-level distribution within the metro, given by 
\begin{equation}
\label{eq:Havg}
\begin{split}
H_{avg}(D) &= \frac{1}{n(V)}\sum_{u\in V}H(\{b_r(u)/b(u)\}) \\
&= -\frac{1}{n(V)}\sum_{u\in V}\sum_{r=1}^{R}\frac{b_r(u)}{b(u)}\log \frac{b_r(u)}{b(u)}, 
\end{split}
\end{equation}
where $H$ is the Shannon entropy. Eq.~\ref{eq:Havg} will take its minimum value of $0$ when the population in $D$ is concentrated entirely into a single category $r$ within each tract, and its maximum value of $\log R$ when all categories have equal representation in each tract within the metro. 

To measure the second feature of interest, the spatial scale of clustering for a metro area, we define the characteristic cluster length scale $\xi(D)$ as 
\begin{equation}
\label{eq:scale}
\xi(D) = \sqrt{\frac{\sum_{k}A(V_k)^2}{A(V)}},    
\end{equation}
where $A(V')$ is the area of tracts in the subset $V'\subseteq V$, and $\mathcal{P}=\{V_k\}_{k=1}^{K}$ is the minimum description length partition of the metro. Eq.~\ref{eq:scale} will take its minimum value of $\sqrt{A(V)/n(V)}$ when each cluster has a spatial extent of $A(V)/n(V)$---the area of a single tract if the tracts were of equal size and each cluster only consisted of a single tract. Conversely, Eq.~\ref{eq:scale} will take its maximum value of $\sqrt{A(V)}$, the length scale of the entire metro, when the data $D$ is best compressed with only a single cluster.   

In Figure \ref{fig:entvcomp} we show how changes in the inverse compression ratio $\eta$ (Eq.~\ref{eq:comp}) correspond to changes in $H_{avg}$ (Eq.~\ref{eq:Havg}) across all 110 metros for each time period in our dataset. In order to account for unobserved heterogeneity in each metro network that is constant in time---for example due to the size and topology of the metro adjacency network---as well as for potentially nonlinear dependencies, ordinary least squares (OLS) regression analysis was performed on the differences in the logarithm of each quantity over each of the periods $1980-1990$, $1990-2000$, and $2000-2010$ (panels (a), (b), and (c) respectively in the figure). All significance results reported in the captions hold up under Bonferroni correction for multiple comparisons \cite{miller1980simult}.

We can see that the inverse compression ratio $\eta$ is in general increasing over all time periods, as the majority of the points in Fig.~\ref{fig:entvcomp} fall above the line $y=0$. The average values of $\eta$ over the four decades are $\{0.74, 0.77, 0.82, 0.85\}$ for $\{1980,1990,2000,2010\}$. In particular, the values of $\eta$ increased substantially between $t=1990$ and $t=2000$, with all metros in our dataset having a positive change in this quantity during this decade. This general pattern of decreasing compressibility, with the greatest change occurring during the $1990-2000$ period, is consistent with the case study analysis in Sec.~\ref{sec:casestudy}.

Looking at Fig.~\ref{fig:entvcomp}, we can also observe a consistently increasing level of tract-level diversity in the metro areas, as illustrated by the majority of points falling to the right of the line $x=0$ in the three plots. The average values of $H_{avg}$ over the four decades are $\{0.57, 0.67, 0.87, 1.02\}$ for $\{1980,1990,2000,2010\}$. This observation is consistent with findings that suburbs have generally become more racially diverse \cite{orfield2013america}, that there are an increasing number of ``no-majority'' communities in which no ethnoracial group makes up more than half of the population \cite{farrell2018no}, and that the diversification of cities in the U.S. is manifested nationwide with no significant regional dependence \cite{dmowska2018spatial}. The Scranton Wilkes-Barre metro area (the rightmost point in Fig.~\ref{fig:entvcomp}) represents a clear outlier regarding changes in overall diversity, as its value of $H_{avg}$ shot up in $2010$, with roughly a $105\%$ increase from relatively low values in the first three decades. The coefficients of determination $r^2$ for the regression analyses reveal that the temporal changes in $H_{avg}$ are highly correlated with the changes in $\eta$ over the same time periods, with the strongest correlation occurring between $2000$ and $2010$. These $r^2$ values, along with the statistically significant p-values of the corresponding regression line slopes (all of which had $p\ll 0.01$), suggest that the small-scale diversity within metros is an important factor for determining the complexity in segregation patterns we see according to Eq.~\ref{eq:comp}.

Indeed, the results in Fig.~\ref{fig:entvcomp} should not be too surprising: Eq.~\ref{eq:Havg} has its origins in the theory of information transmission and can itself be used as a measure of spatial segregation \cite{wright2014patterns}, like the compressibility in Eq.~\ref{eq:comp}. However, Eq.~\ref{eq:Havg} accounts only for diversity at small spatial scales, while the compressibility in Eq.~\ref{eq:comp} accounts for both small-scale diversity as well as large scale homogeneity within clusters. In this way, both large-scale segregation and small-scale diversity will affect the compressibility, and therefore we need to examine both factors to determine which is a more dominant force associated with the increasing complexity we see in metros according to Eq.~\ref{eq:comp}.

As shown in Fig.~\ref{fig:scalevcomp}, however, we observe no clear trend in the changes in the characteristic cluster length scales $\xi$ (Eq~\ref{eq:scale}) across metros for each time period, with roughly half of the metros in each time period having decreasing values $\xi$, and half having increasing values of $\xi$. The metros that comprise these two halves also differ across time periods: only $18$ of the $110$ metros studied had monotonically increasing or decreasing values of $\xi$ across all time periods (compared to $105$ of $110$ metros having a value of $H_{avg}$ that increased throughout all decades). The $r^2$ values for the regression analyses in Fig.~\ref{fig:scalevcomp} indicate that the temporal changes in $\xi$ are poorly correlated with the changes in $\eta$ over the same time periods, with $r^2$ values in two of the decades even rounding to $0$ up to two decimal places. The p-values corresponding to the slopes of the regression lines plotted do not indicate any statistically significant linear relationship between the plotted variables---$p=0.52$, $0.13$, and $0.89$ for panels (a), (b), and (c) respectively. These results suggest that the increasing complexity of segregation patterns we observe across metros is not substantially affected by the characteristic spatial scale at which the units can be optimally clustered in each metro (at least when considering census tracts as the fundamental unit, which obscures segregation patterns at smaller scales \cite{krupka2007big,dmowska2021improving}).  

An important additional consideration to take into account is the effect of population, as the population in most of the metros is increasing over time, and it is reasonable to expect that this may affect the compressibility of the data. In Fig.~\ref{fig:popanalyses} we plot, in the same style as Fig~\ref{fig:entvcomp}, the changes in population and the changes in compressibility of the metros over time. We can see from the OLS $r^2$ values that there is very little to no dependence between the population changes and the changes in compressibility of the metros (consistent with the discussion in Sec.~\ref{sec:optimization}). We can also consider the effects of population and average diversity simultaneously through the following regression with city-level fixed effects
\begin{equation}
\log \eta_{ct} = \beta_1 \log H_{avg,ct} + \beta_2 \log b_{ct} + \alpha_{c} + \epsilon_{ct},   
\end{equation}
where $c$ and $t$ index metros and decades respectively, $\beta_{1,2}$ are regression coefficients, $\alpha_c$ is an unobserved time-invariant source of heterogeneity specific to metro $c$ (for example, based on metro $c$'s adjacency network topology), and $\epsilon$ is a noise term. We can then run a regression for the first differences estimators to remove the heterogeneity $\alpha_c$ and identify the effect $\log H_{avg}$ and $\log b$ have on $\log \eta$ when considered together. By partitioning the individual contributions of each term to the variance in the dependent variable  \cite{gromping2007relative}, we find a relative importance of $97.8\%$ for $\log H_{avg}$ versus only $2.1\%$ for $\log b$, indicating that the average local diversity is a much more important factor for determining the compressibility than population. 
        
Altogether, this analysis indicates that segregation patterns in large U.S. metros are becoming more complex over time from the perspective of information compression. The small-scale diversification of these metros plays an important role in increasing the complexity of these segregation patterns, while changes in population and large-scale spatial clustering among ethnoracial groups are likely not major contributors.


\section{Conclusions}

Here we have presented a network regionalization algorithm based on the minimum description length principle for partitioning a set of spatial units with distributional metadata into contiguous clusters. Our method requires no user input, learning the natural clusters that result in a maximally compressed representation of the data. We demonstrate that our approach can effectively recover synthetically planted clusters in noisy spatial data and that it returns a partitioning of ethnoracial census data in U.S. metropolitan areas that can allow for insights about the ethnoracial segregation patterns in these metros. We find that the segregation patterns in these metros have become increasingly complex over time, in part due to the increasing small-scale ethnoracial diversity of the metros over the time period studied. 

There are a number of ways our method can be extended in future work. Our current formulation requires the spatial data of interest to take the form of a single discrete set of counts within each unit, but it may be possible to perform a similar description length calculation for the transmission of multiple spatial covariates simultaneously by employing the combinatorial form of the shared information between these covariates and transmitting a contingency table indexed by groups of covariates rather than a single covariate (similar in spirit to the encoding in \cite{NCY20}). One could also develop objectives for clustering with ordinal or continuous metadata by considering the transmission on a per-symbol basis and using continuous approximations for the entropy and mutual information. This would allow us to perform regionalization with respect to a variety of attributes of interest with variable data types, for example race and income, all at once. Extension of our transmission procedure to a multi-step, hierarchical encoding scheme may also prove useful, as this would allow for multiscalar regionalization. It is also possible to include additional penalties in the regionalization objective function we use in the form of Lagrange multipliers that enforce constraints on the size, shape, or populations of the clusters, which may make our method more suitable for policy-driven applications of regionalization. Additionally, using description length-based data imputation \cite{vreeken2008filling} one may be able to adapt our method to be robust for use with incomplete data. Finally, a comprehensive numerical comparison between the method of this paper and existing regionalization methods would shed light on the advantages and disadvantages of the MDL approach to regionalization (see Appendix~\ref{sec:methodcomparison} for a qualitative comparison with similar existing methods).


\bigskip\noindent\textbf{Acknowledgments:} The author thanks Mark Newman for useful discussions regarding the use of mutual information for spatial clustering, and Phil Chodrow for useful discussions about the data. This work was funded by the US Department of Defense NDSEG fellowship program.

\bigskip\noindent\textbf{Data availability:} 
All data needed to evaluate the conclusions in the paper are present in the paper or are available at \url{https://github.com/aleckirkley/MDL_regionalization}.

\bigskip\noindent\textbf{Code availability:} 
The regionalization algorithm presented in this paper is available at \url{https://github.com/aleckirkley/MDL_regionalization}.

\bigskip\noindent\textbf{Competing interests:} 
The authors declare that there are no competing interests.


\clearpage
\appendix

\section{Algorithm time complexity}\label{sec:runtime}

\begin{figure}[h]
    \centering
    \includegraphics[width=1\columnwidth]{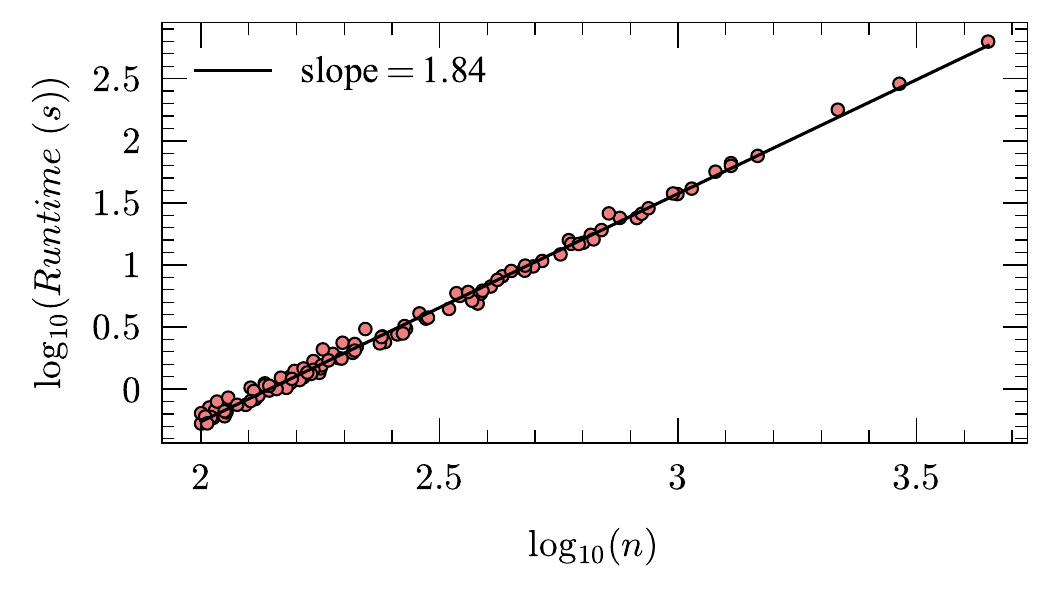}
    \caption{Log total run time (in seconds) versus the log number of nodes in the network for all cities studied in the main text for 2010, showing a scaling of $(\text{Runtime})\sim \Ord(n(V)^{1.84}) < \Ord(n(V)^{2})$. Only the 2010 results are shown, but all decades presented nearly identical results as the network topology remained unchanged and the final number of clusters did not change significantly from decade to decade.}
    \label{fig:runtime}
\end{figure}

In Fig.~\ref{fig:runtime} we plot the log total run time (in seconds) of the MDL regionalization algorithm versus the log number of nodes in the network for all cities studied in Sec.~\ref{sec:metros} in the main text, for 2010 (the results are nearly identical for all decades). We can see a scaling of $(\text{Runtime})\sim \Ord(n(V)^{1.84}) < \Ord(n(V)^{2})$ for the full clustering procedure. 

\section{Comparison with exact solution}\label{sec:exactcomparison}

To compare our greedy approach with the results from exact enumeration, we perform the following experiment. For each trial, we choose a metro at random from the set analyzed in Sec.~\ref{sec:metros}, and select a node from this network at random. The full metro itself cannot be analyzed with exhaustive enumeration, so we then perform a random walk of a specified length $n$ and take the resulting induced subgraph of size $n$ within the metro, which is guaranteed to be spatially contiguous. Very small subgraphs are necessary in order to completely enumerate the partitions of the network for the exact algorithm, and so we set $n=12$ for the experiments.  Finally, both the exact algorithm and greedy algorithm are evaluated on the resulting data, and the compression ratios (Eq.~\ref{eq:DL}) for each algorithm are computed for comparison. 

\begin{figure}[h]
    \centering
    \includegraphics[width=1\columnwidth]{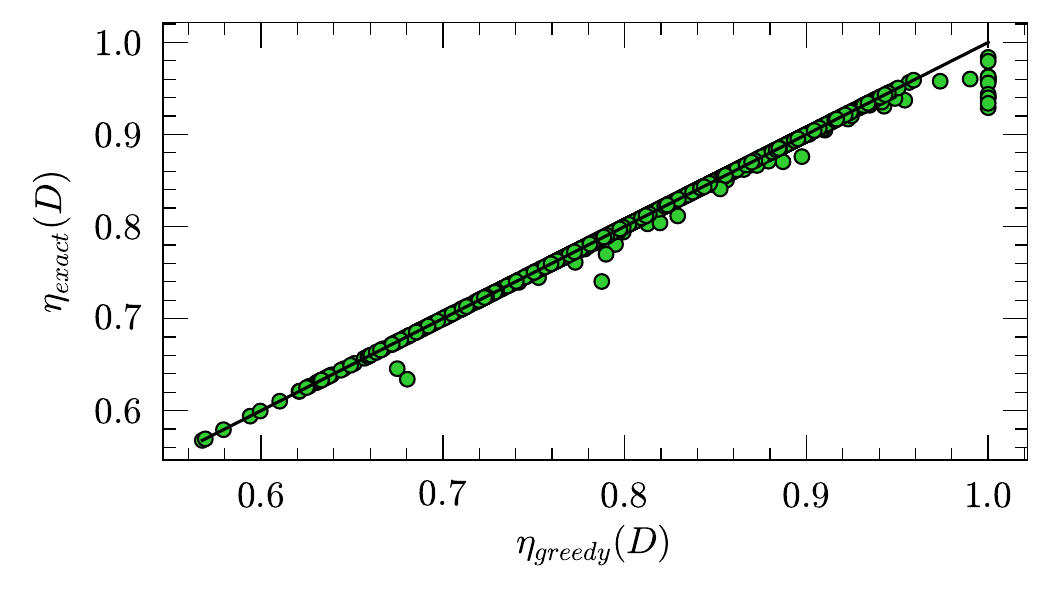}
    \caption{Comparison of greedy and exact results for minimizing the description length objective in Eq.~\ref{eq:DL}. The compression ratio (Eq.~\ref{eq:comp}) for the exact and greedy methods is plotted for $1000$ simulated network datasets using the procedure described above. The line of equality is plotted for reference in black.}
    \label{fig:exactcomparison}
\end{figure}

The experiment was run for $1000$ iterations and the results are plotted in Fig.~\ref{fig:exactcomparison}. We can see that the greedy and exact solutions in general tend to give nearly identical compression ratios for most of the simulated datasets, with a few exceptions where the exact algorithm performs noticeably better. This similarity in compression performance persists over simulated datasets with a wide range of levels of compressibility.

\begin{figure}
    \centering
    \includegraphics[width=1\columnwidth]{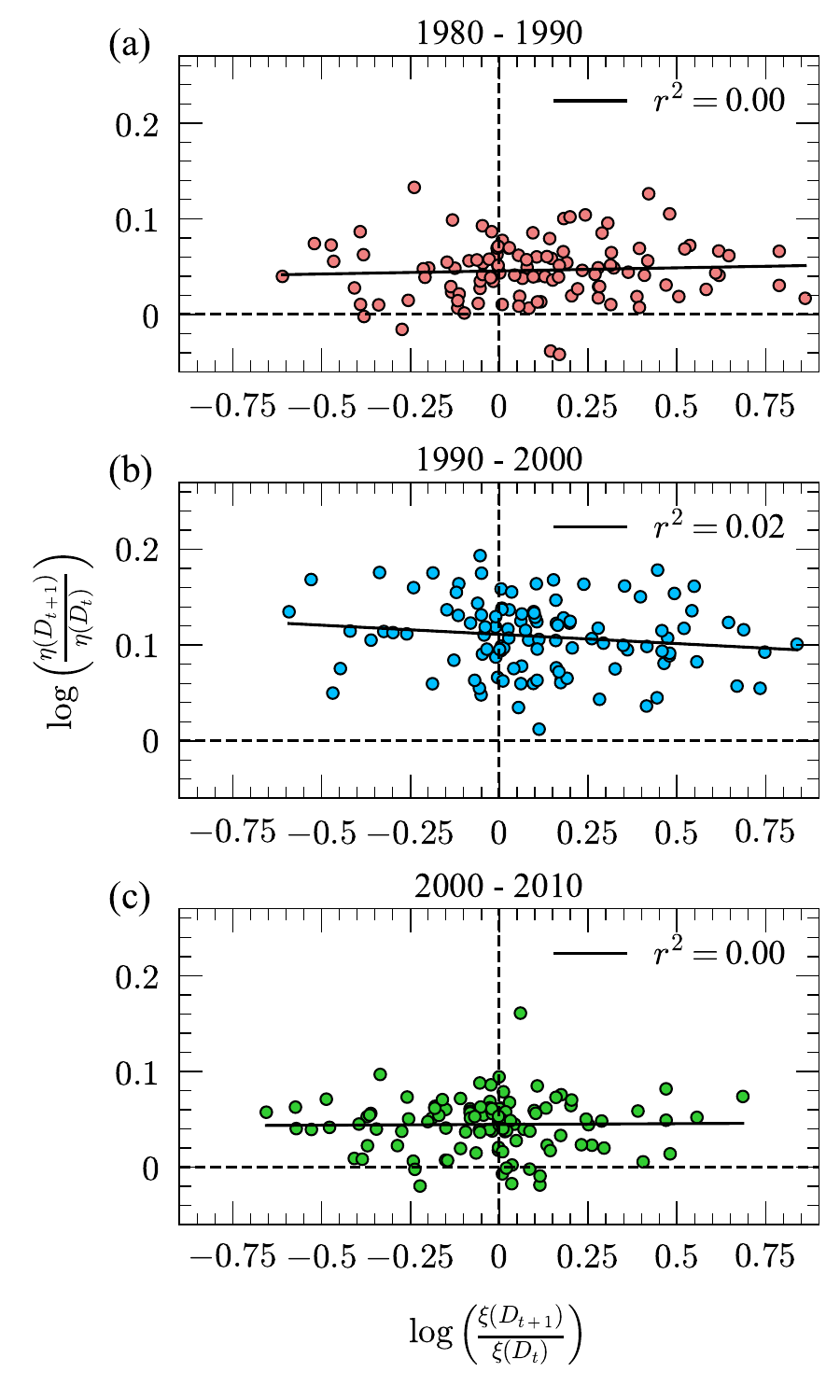}
    \caption{Log-ratio of consecutive inverse compression ratios $\eta$ (Eq.~\ref{eq:comp}) versus the log-ratio of consecutive cluster scales $\xi$ (Eq.~\ref{eq:scale}), in U.S. metros over the decades spanning $t=1980$ to $t=2010$. None of the regression line slopes were statistically significant, even at the $0.10$ significance level. Without grouping the changes by decade, we find $r^2=0.01$, and that the slope is again not significant at the $0.10$ significance level.}
    \label{fig:scalevcomp}
\end{figure}

\section{Effect of population heterogeneity}\label{sec:populationhet}

\begin{figure}[h]
    \centering
    \includegraphics[width=1\columnwidth]{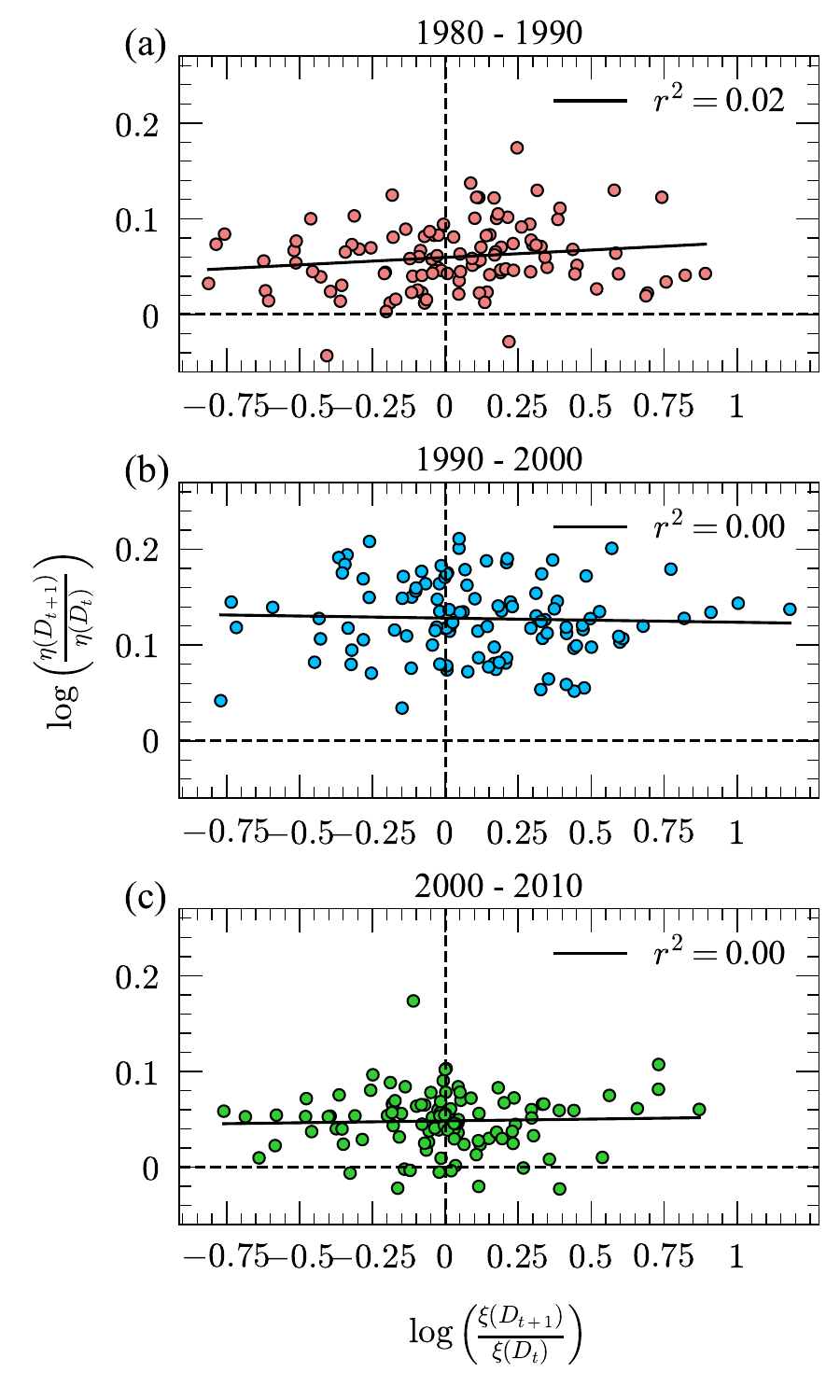}
    \caption{Re-creation of Fig.~\ref{fig:scalevcomp} with all tract-level populations set to the metro-wide average for each city/decade, also presenting similar results.}
    \label{fig:equalcomparisonscale}
\end{figure}

\begin{figure}[h]
    \centering
    \includegraphics[width=\columnwidth]{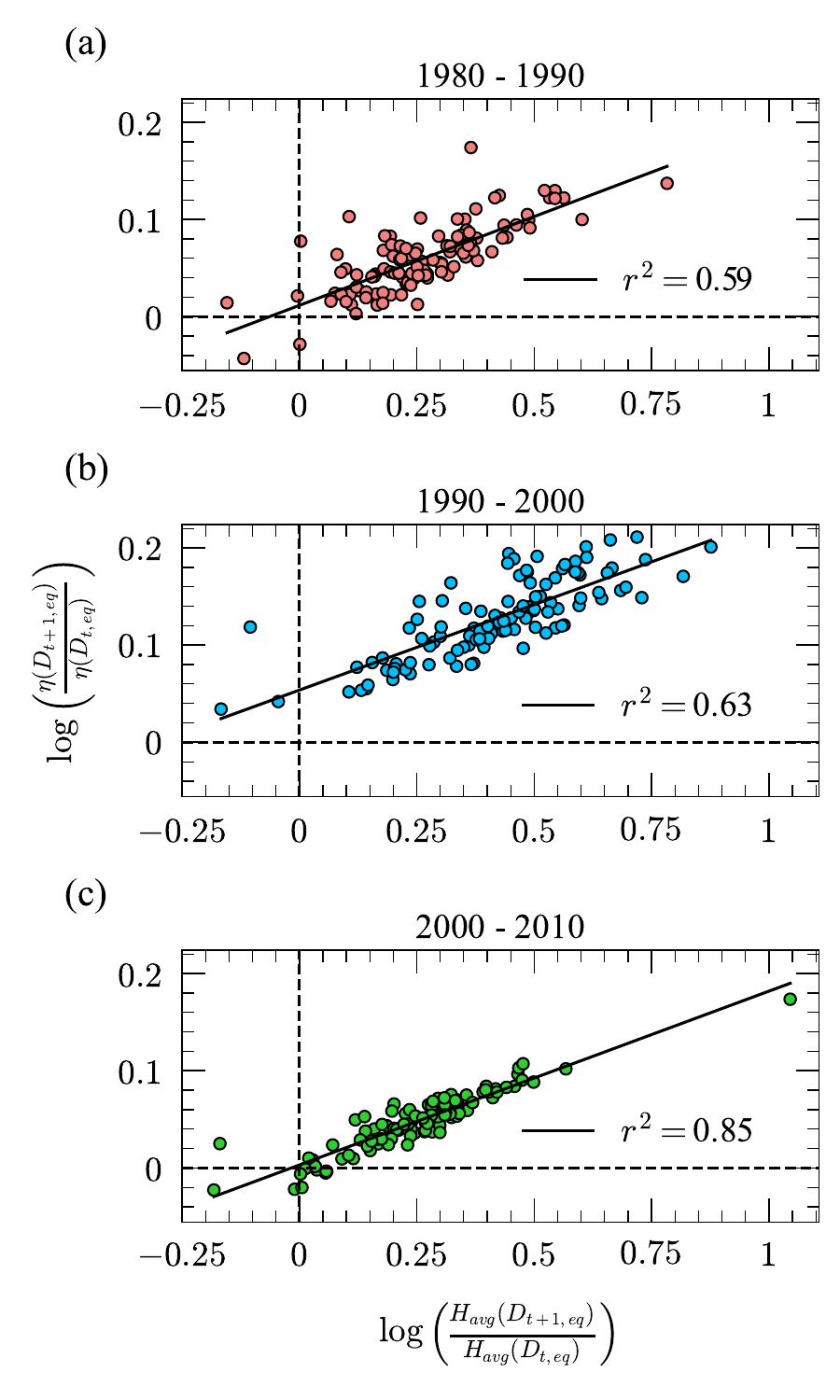}
    \caption{Re-creation of Fig.~\ref{fig:entvcomp} with all tract-level populations set to the metro-wide average for each city/decade. We can see little qualitative difference in the findings, with strong correlations across all decades and only slight changes in compressibilities.}
    \label{fig:equalcomparisonavgH}
\end{figure}

In Fig.~\ref{fig:equalcomparisonavgH}, we compare the compression results obtained for all the metros analyzed in Sec.~\ref{sec:metros}, before and after setting all the tract-level populations equal to the city-wide average population. We can see that the correlation between compressibility (Eq.~\ref{eq:comp}) and average entropy (Eq.~\ref{eq:Havg}) for all cities is qualitatively similar with and without population heterogeneity across tracts. In Fig.~\ref{fig:equalcomparisonscale}, we repeat this experiment for compressibility versus cluster scale (Eq.~\ref{eq:scale}), finding again a high level of similarity in the results.

\begin{figure}[h]
    \centering
    \includegraphics[width=\columnwidth]{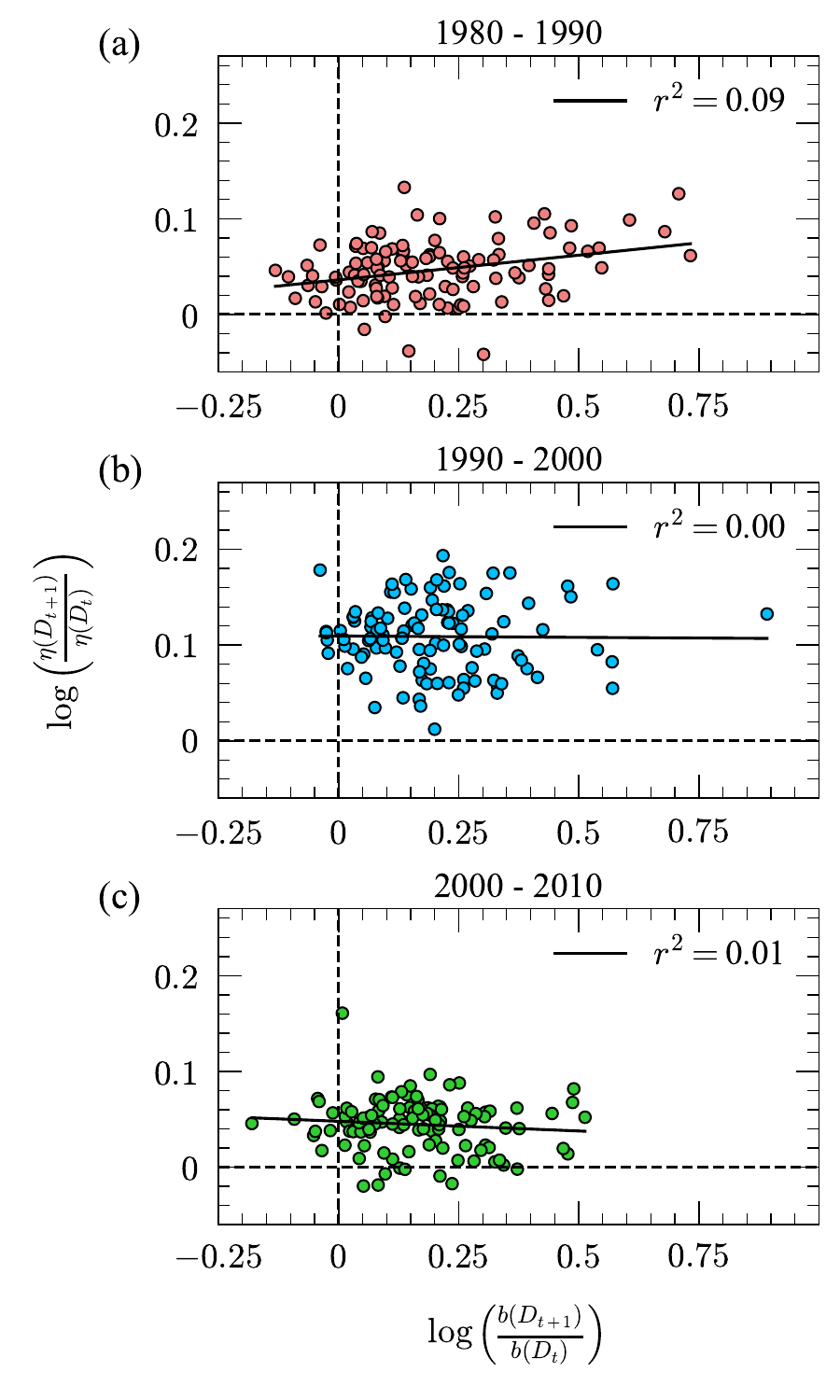}
    \caption{Assessing the effects of population. Log of consecutive compression ratios versus log of consecutive populations for all cities over (a) 1980 - 1990, (b) 1990 - 2000, and (c) 2000 - 2010.}
    \label{fig:popanalyses}
\end{figure}

\begin{figure*}
    \centering
    \includegraphics[width=0.9\textwidth]{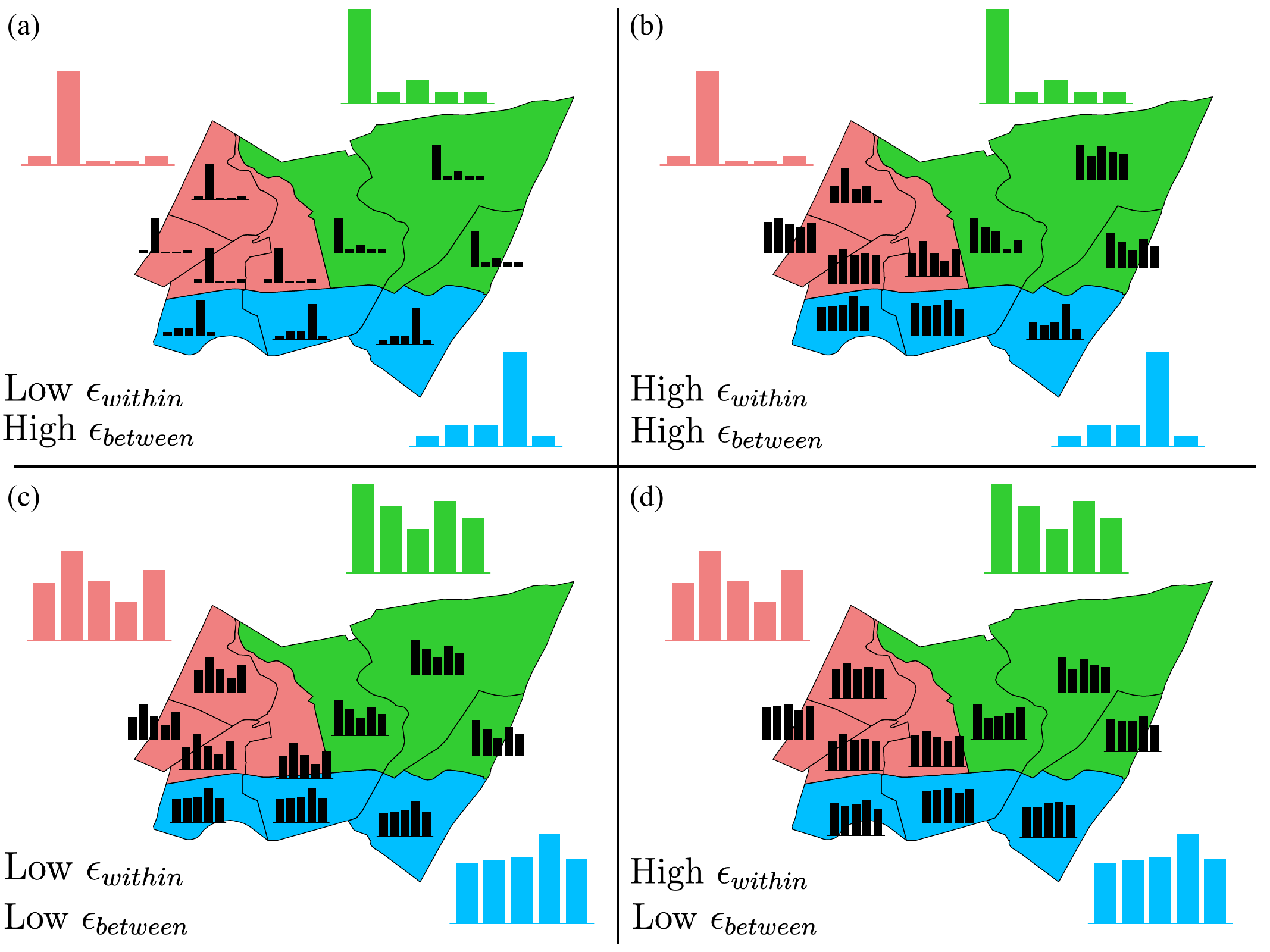}
    \caption{Example realizations of the model in Sec.~\ref{sec:synthetic}. The within-cluster noise $\epsilon_{within}$ tunes the level of variability in the tract-level distributions (black) within each cluster from the latent distribution assigned to the cluster (red/blue/green), with higher values corresponding to greater levels of noise. The between-cluster noise $\epsilon_{between}$ tunes the level of variability in the latent cluster-level distributions, with higher values corresponding to greater levels of variability. Cluster recovery is easiest with low $\epsilon_{within}$ and high $\epsilon_{between}$ (panel (a)), as there are high levels of distribution homogeneity within clusters and high levels of heterogeneity across clusters. Cluster recovery is hardest with high $\epsilon_{within}$ and low $\epsilon_{between}$ (panel (d)), as there are low levels of distribution homogeneity within clusters and low levels of heterogeneity across clusters.}
    \label{fig:syntheticdiagram}
\end{figure*}

\section{Comparison with existing methods}\label{sec:methodcomparison}

An apples-to-apples numerical performance comparison between the regionalization method presented in this paper and existing regionalization methods is difficult, since existing methods require as input an unspecified dissimilarity function, a desired final threshold for some function of the partition such as average population, or the number of clusters (all of which will affect the final partition) and there is no a priori way to choose these inputs to the algorithm. However, we can sensibly compare these algorithms with our own in a more qualitative way by considering the nature of how these algorithms perform optimization, their computational complexity, and the free input parameters they require. Given the great variety of regionalization methods that have been proposed in previous research, we ground our discussion by considering methods that are deterministic, do not explicitly constrain the size/shape of clusters, and can take as input categorical data distributions across a set of spatial units. These methods are most directly comparable to the method presented in this paper. A summary of the considered set of regionalization algorithms is shown in Table~\ref{table:comparison}.

The first method we consider is the SKATER algorithm \cite{assunccao2006efficient}, which takes as free parameters a dissimilarity function between spatial units and a termination criterion (e.g. the desired number of clusters), and aims to minimize the sum of dissimilarities between the units in each cluster and their corresponding cluster-level averages until the termination criterion is met. The SKATER algorithm first constructs the minimum spanning tree on the network of adjacent spatial units according to the specified dissimilarity function, then repeatedly removes the edge from the tree such that the resulting connected clusters after this edge removal minimize the cluster-level objective. 

The second method we consider is the REDCAP algorithm \cite{guo2008regionalization}, which is inspired by the SKATER algorithm and actually encompasses a collection of single-linkage, average-linkage, and complete-linkage spatially-constrained hierarchical clustering methods. These methods consider the repeated merging of regions with the lowest level of dissimilarity according to the linkage criterion that is chosen, and can be adapted to only consider bordering units (``first-order constraint'') or all constituent units in each cluster (``full-order constraint''). The variation of the algorithm that was found to perform best in a comparative setting was the full-order, complete linkage clustering \cite{aydin2021quantitative}, and so this is the method for which we report the summary in Table~\ref{table:comparison}. 

The last two methods we consider are those proposed in \cite{chodrow2017structure}, which consider the information loss associated with the merging of regions according to the class of Bregman Divergences, the particular choice of divergence needing specification prior to running the algorithm. The agglomerative greedy algorithm proposed in \cite{chodrow2017structure} repeatedly merges the regions with the smallest Bregman divergence until the desired number of clusters is reached. In principle this algorithm should be possible to implement in a very similar manner to the greedy algorithm presented in this paper, since cluster-level objectives are decoupled, and so it should also run in $\Ord(n^\alpha)$ time with $\alpha \in (1,2)$. The specific exponent $\alpha$ may vary depending on the sizes and shapes of clusters output by the algorithm. 

A refinement to this method is also proposed in the paper, which first maps the Bregman divergence to a similarity measure between each pair of spatial units through a Gaussian kernel and then performs a k-means clustering using the eigenspaces of the resulting graph Laplacian. After this spectral preprocessing step is performed to obtain intermediate clusters, the greedy agglomerative method is run until the desired final number of regions is reached. Using standard eigendecomposition methods such as the power method, spectral partitioning of this kind can be run with $\Ord(n^2)$ time complexity, assuming the number of eigenvectors used does not scale with the system size \cite{Newman06c}.   

We can see from Table~\ref{table:comparison} that all methods being compared have a greedy and/or agglomerative nature to their optimization, indicating that this is a common means for constructing scalable non-stochastic approximate solutions to regionalization problems in two dimensions. (In one dimension, a large class of regionalization objectives can be solved exactly in polynomial time using dynamic programming \cite{wang2011ckmeans}.) We can also see that the method presented here is the only one which does not require any free parameters as input, which presents a major advantage for users who do not want to impose any assumptions about the clustering structure of the data they are regionalizing. Finally, we can see that all algorithms in this group have roughly the same time complexity, with the method in this paper and the agglomerative algorithm of \cite{chodrow2017structure} having a slight edge with sub-quadratic time complexities.

\begin{table*}
\begin{tabular}{|p{4cm}|p{4cm}|p{4cm}|p{4cm}|}
\hline 

 \textbf{Algorithm} & \textbf{Optimization} & \textbf{Free parameters} & \textbf{Time complexity} \\\hline  
 
 MDL (this paper) & Greedy & None & $ \approx\Ord(n^{\alpha}),~~\alpha\in (1,2)$ \\\hline
 
 SKATER \cite{assunccao2006efficient} & Greedy & dissimilarity function, termination criterion  & $\Ord(n^2\log n)$ \\\hline
 
 REDCAP \cite{guo2008regionalization} & Agglomerative hierarchical & dissimilarity function, order criterion   & $\Ord(n^2\log n)$ \\\hline
 
 Agglomerative Bregman \cite{chodrow2017structure} & Greedy & choice of Bregman divergence, number of clusters  & $ \approx\Ord(n^{\alpha}),~~\alpha\in (1,2)$ \\\hline
 
 Spectral Bregman \cite{chodrow2017structure} & Spectral, greedy & choice of Bregman divergence, number of clusters, scale parameter & $\Ord(n^2)$ \\\hline

\end{tabular}

\caption{Summary of characteristics for a number of deterministic regionalization algorithms. Computational complexity is assessed in terms of the total number of spatial units $n$ being partitioned.
\label{table:comparison}
}
\end{table*}

\end{document}